\documentclass[preprint]{article}
\usepackage[latin9]{inputenc}
\usepackage[a4paper]{geometry}
\usepackage{color}
\usepackage{array}
\usepackage{float}
\usepackage{amsmath}
\usepackage{graphicx}
\usepackage{setspace}

\makeatletter

\providecommand{\tabularnewline}{\\}

\newcommand{\lyxaddress}[1]{
\par {\raggedright #1
\vspace{1.4em}
\noindent\par}
}

\usepackage{times}
\usepackage{txfonts}

\makeatother

\begin{document}

\title{Communicability Angles Reveal Critical Edges for Network Consensus
Dynamics }

\author{Ernesto Estrada$^{1}$, Eusebio Vargas-Estrada$^{1}$, Hiroyasu Ando$^{2}$}

\maketitle

\lyxaddress{\begin{center}
$^{1}$Department of Mathematics and Statistics, University of Strathclyde,
26 Richmond Street, Glasgow G1 1HX, United Kingdom 
\par\end{center}}

\lyxaddress{\begin{center}
$^{2}$Division of Policy and Planning Sciences, Faculty of Engineering,
Information and Systems, University of Tsukuba\\
 1-1-1 Ten-noudai, Tsukuba, 305-8573 Japan
\par\end{center}}
\begin{abstract}
We consider the question of determining how the topological structure
influences a consensus dynamical processes taking place on a network.
By considering a large dataset of real-world networks we first determine
that the removal of edges according to their communicability angle---an
angle between position vectors of the nodes in an Euclidean communicability
space---increases the average time of consensus by a factor of $5.68$
in real-world networks. The edge betweenness centrality also identifies---in
a smaller proportion---those critical edges for the consensus dynamics,
i.e., its removal increases the time of consensus by a factor of $3.70$.
We justify theoretically these findings on the basis of the role played
by the algebraic connectivity and the isoperimetric number of networks
on the dynamical process studied, and their connections with the properties
mentioned before. Finally, we study the role played by global topological
parameters of networks on the consensus dynamics. We determine that
the network density and the average distance-sum---an analogous of
the node degree for shortest-path distances, account for more than
80\% of the variance of the average time of consensus in the real-world
networks studied.
\end{abstract}
PACS number(s): 89.75.-k, 02.10.Ox, 05.45.Xt

\section{Introduction}

Complex networks are ubiquitous in many real-world systems ranging
from biological and ecological to social and infrastructural ones
\cite{Estrada_book}. One of the most important aspects of these networked
systems is the transmission of information from one node to another.
It can be argued indeed that networks exist for facilitating the information
transmission in those complex systems. The nodes of these networks
represent the entities of the complex system and their connections
represent the interactions among these entities from which information
flows from one node to another. Information is understood here generically
and can represent such a variety of things like the transfer of material
or energy to the spread of diseases or rumors.

The diffusion of information over a network is usually analyzed by
considering that the state of the nodes of the graph at time $t$
are stored in a vector $\vec{u}\left(t\right)$. Then, the variation
of the state of the node $i$ with time is controlled by the equation
\cite{Mesbahi_Egerstedt_Book,Olfati-Saber,Ren-Beard-Atkins}:

\begin{equation}
\vec{\dot{u_{i}}}\left(t\right)=\sum_{\left(i,j\right)\in E}\left(\vec{u_{j}}\left(t\right)-\vec{u_{i}}\left(t\right)\right),\ i=1,2,\ldots,n,\label{eq:consensus node}
\end{equation}

where the sum is taken over all pairs of connected nodes in the network.
This simple diffusion model is typically used for analyzing consensus
dynamics on networks, in which the pairs of connected nodes of the
graph try to reach an agreement over a given topic, e.g., opinions,
position in space, etc., and the network as a whole collapses to a
steady state of consensus. Consensus protocols, as they are known
in technological applications, \textcolor{black}{have been widely
used in the study of }wireless sensor networks (WSNs) and peer-to-peer
networks, where the problem consists of making the scalar states of
a set of agents converge to the same value under local communication
constraints \cite{WSN_1,WSN_2,Consensus in distributed systems_1,Consensus in distributed systems_2}.
In social network analysis the consensus dynamics plays a fundamental
role in understanding the dynamics of information spreading among
actors in a social system and it has been applied for a diverse series
of real-world situations \cite{Social consensus,Social consensus_2,Social consensus_3,Social consensus_4}.

A very important question when analyzing a dynamical model, like consensus,
is to understand the role played by the network structure on the dynamical
process. These structure-dynamics relations allow us to understand
what are the roles played by different structural parameters over
the dynamics, which permit to engineering the systems to change their
dynamical properties. The important problem of network controllability
\cite{Network controllability_1,Network controllability_2,Network controllability_3,Network controllability_4},
for instance, very much resides in understanding the influence of
structural parameters on the control of a dynamical process taking
place on the network. Here, we explore the structure-dynamics relations
for the consensus model in real-world networks. First, we consider
the problem of identifying critical edges for the consensus dynamics,
i.e., those edges whose removal significantly increase the average
time of consensus in the network. We found that among a few structural
parameters describing the capacity of an edge to transmit information
through it, the communicability angle identifies the most critical
edges for the consensus dynamics in a wide variety of networks. We
then consider the influence of a few structural parameters characterizing
global structural properties of networks over the average time of
consensus. We find that the network density and the average shortest
path distance are global indicators of the network capacity to perform
consensus in an efficient way.

\section{Preliminaries}

Here we represent networks by means of simple graphs. A \textit{graph}
$\varGamma=(V,E)$ is defined by a set of $n$ nodes (vertices) $V$
and a set of $m$ edges $E=\{(u,v)|u,v\in V\}$ between the nodes.
The graph is said to be \textit{undirected} if the edges are formed
by unordered pairs of vertices. A \textit{path} of length $k$ in
$G$ is a set of nodes $i_{1},i_{2},\ldots,i_{k},i_{k+1}$ such that
for all $1\leq l\leq k$, $(i_{l},i_{l+1})\in E$ , and there are
no repeated nodes. The graph is \textit{connected} if there is a path
connecting every pair of nodes. The length of the shortest of all
paths connecting two nodes in the graph is known as the\textit{ shortest
path distance} between the corresponding nodes. A graph with unweighted
edges, no self-loops (edges from a node to itself), and no multiple
edges is said to be \textit{simple}. Hereafter we will always consider
undirected, simple, and connected networks.

The matrix $A=\left(a_{uv}\right)$, called the \textit{adjacency
matrix} of the graph, has entries 
\[
a_{uv}=\left\{ \begin{array}{ll}
1 & \mbox{if }(u,v)\in E\\
0 & \mbox{otherwise}
\end{array}\right.\qquad\forall u,v\in V.
\]

The adjacency matrix can be decomposed by $A=Q\Lambda Q^{T}$, with
$\Lambda$ a diagonal matrix containing the eigenvalues of $A$ and
$Q=[\mathbf{q}_{1},\ldots,\mathbf{q}_{n}]$ an orthogonal matrix containing
the associated eigenvectors.

The degree $k_{i}$ of the node $i$ is the number of edges incident
to it, equivalently $k_{i}=\sum_{j}a_{ij}$. We will designate by
$\delta=\min\left(k_{i}\right)$ and $\varDelta=\max\left(k_{i}\right)$
the minimum and maximum degree in the network. The matrix $K=diag\left(k_{i}\right)$
is named the degree matrix of the graph. The matrix $\mathcal{L}=K-A$
is known as the graph Laplacian. It has entries \cite{Mesbahi_Egerstedt_Book,Olfati-Saber}

\[
\mathcal{L}_{uv}=\left\{ \begin{array}{ll}
k_{i} & \mbox{if }u=v\\
-1 & \mbox{if }\left(u,v\right)\in E\\
0 & \mbox{otherwise}
\end{array}\right.\qquad\forall u,v\in V.
\]

The Laplacian matrix is positive semi-definite with eigenvalues denoted
here by: $0=\mu_{1}\leq\mu_{2}\leq\cdots\leq\mu_{n}$. If the network
is connected the multiplicity of the zero eigenvalue is equal to one,
i.e., $0=\mu_{1}<\mu_{2}\leq\cdots\leq\mu_{n}$ and the smallest nontrivial
eigenvalue $\mu_{2}$ is known as the \textit{algebraic connectivity}
of the graph \cite{Fiedler,algebraic connectivity}. Let $U$ be the
matrix of orthonormalized eigenvectors $\vec{\psi}_{j}$ of $\mathcal{L}$,
i.e., $V=\left[\begin{array}{ccc}
\vec{\psi}_{1} & \cdots & \vec{\psi}_{n}\end{array}\right]$. The eigenvector $\vec{\psi}_{2}$ associated with the algebraic
connectivity is known as the Fiedler vector \cite{Fiedler}. Let $\varSigma$
be the diagonal matrix of eigenvalues of the Laplacian matrix. Then,
$\mathcal{L}=V\varSigma V^{T}$.

\section{Consensus Dynamics}

The consensus dynamics equation (\ref{eq:consensus node}) can be
written as follows for the kind of graphs we analyze in this work
\begin{equation}
\vec{\dot{u_{i}}}\left(t\right)=-\sum_{j=1}^{n}a_{ij}\left(\vec{u_{i}}\left(t\right)-\vec{u_{j}}\left(t\right)\right),\ i=1,2,\ldots,n.
\end{equation}

This equation indicates that the evolution of the state of the node
$i$ in time depends on the 'agreement' that this node reaches with
all its nearest neighbors. It is obvious now that we can write (\ref{eq:consensus node})
by using the Laplacian matrix of the graph:

\begin{onehalfspace}
\begin{eqnarray}
\vec{\dot{u}}\left(t\right) & = & -\mathcal{L}\vec{u}\left(t\right),\label{eq:consensus model}\\
\vec{u}\left(0\right) & = & \vec{u}_{0}.
\end{eqnarray}

The solution of this equation is:

\begin{equation}
\vec{u}_{t}=e^{-t\mathcal{L}}\vec{u}_{0}.
\end{equation}

where $0=\mu_{1}<\mu_{2}\leq\cdots\leq\mu_{n}$ are the eigenvalues
and $\vec{\psi}_{j}\left(p\right)$ is the $p$th entry of the corresponding
$j$th eigenvector of the Laplacian matrix. Then, the solution of
the consensus equation on the graph is given by

\begin{equation}
\vec{u_{t}}=e^{-t\mu_{1}}\left(\vec{\psi}_{1}\cdot\vec{u}_{0}\right)\vec{\psi}_{1}+e^{-t\mu_{2}}\left(\vec{\psi}_{2}\cdot\vec{u}_{0}\right)\vec{\psi}_{2}+\cdots+e^{-t\mu_{n}}\left(\vec{\psi}_{n}\cdot\vec{u}_{0}\right)\vec{\psi}_{n},\label{eq:spectral consensus}
\end{equation}

\end{onehalfspace}

where $\vec{x}\cdot\vec{y}$ represents the inner product of the corresponding
vectors. When the time tends to infinity every node tends to the state
dictated by the average of the values of the initial condition. This
state is usually known as the \textit{consensus set} \cite{Mesbahi_Egerstedt_Book}
and it can be formally defined as the set $\mathcal{A}\subseteq\mathbb{R}^{n}$
which is the subspace $span\left\{ 1\right\} ,$ i.e.,

\begin{equation}
\mathcal{A}=\left\{ \vec{u}\in\mathbb{R}^{n}\left|\vec{u_{i}}=\vec{u_{j}},\ \forall i,j\in V\right.\right\} .
\end{equation}
The following is a well-known result in the theory of consensus dynamics
on networks.

Let $G$ be a connected graph. Then, the consensus dynamics converges
to the agreement set with a rate of convergence that is dictated by
$\mu_{2}$. That is, as $t\rightarrow\infty$

\begin{equation}
\vec{u}_{t}\rightarrow\left(\vec{\psi}_{1}\cdot\vec{u}_{0}\right)\vec{\psi}_{1}=\dfrac{\vec{1}\vec{\cdot u}_{0}}{n}\vec{1}
\end{equation}

and hence $\vec{u}_{t}\rightarrow\mathcal{A}$. As $\mu_{2}$ is the
smallest positive eigenvalue of the graph Laplacian, it dictates the
slowest mode of convergence in the equation (\ref{eq:spectral consensus}).

For the sake of simulations it is sometimes useful to consider the
discrete-time model of consensus, whose equation can be written as
follows \cite{Mesbahi_Egerstedt_Book,Olfati-Saber}:

\begin{equation}
u_{i}\left(k+1\right)=u_{i}\left(k\right)+\epsilon\sum_{j=1}^{n}a_{ij}\left(u_{j}\left(k\right)-u_{i}\left(k\right)\right),\label{eq:discrete}
\end{equation}

where $0<\epsilon<k_{max}^{-1}$ is the time step for the simulation.
The equation \ref{eq:discrete} can be written in matrix form as follows:

\begin{equation}
\vec{u}\left(k+1\right)=\left(I-\epsilon L\right)\vec{u}\left(k\right),\label{eq:discrete time}
\end{equation}

where $I$ is the identity matrix. The matrix $\left(I-\epsilon L\right)$
is usually known as the Perron matrix \cite{Olfati-Saber}.

\section{Time of Consensus in Networks }

As we have seen in the Section 3 the consensus dynamics is controlled
by the Laplacian matrix of the network. Here we are interested in
considering the influence of the network structure, as captured by
the spectral properties of the network Laplacian, on the time of consensus
$t_{c}$, i.e., the time for which $\left|\vec{u_{i}}-\vec{u_{j}}\right|\leq\delta$,
where $\delta$ is a given threshold. First, we write the eq. (\ref{eq:spectral consensus})
for a given node $p$ as

\begin{onehalfspace}
\begin{equation}
\vec{u}_{t}\left(p\right)=\sum_{q=1}^{n}\vec{u}_{0}\left(q\right)\sum_{j=1}^{n}\vec{\psi}_{j}\left(p\right)\vec{\psi}_{j}\left(q\right)e^{-t\mu_{j}},\label{eq:spectral one node}
\end{equation}

\end{onehalfspace}

which represents the evolution of the state of the corresponding node
as time evolves. Now, let us consider that the time tends to the time
of consensus $t\rightarrow t_{c}$, where $t_{c}$ is the time at
which $u_{t}\rightarrow\left(\vec{\psi}_{1}^{T}\vec{u}_{0}\right)\vec{\psi}_{1}$.
Let us designate this time by $t_{c}^{-}$

\begin{onehalfspace}
\begin{equation}
\vec{u}_{t_{c}^{-}}\left(p\right)=\frac{1}{n}\sum_{q=1}^{n}\vec{u}_{0}\left(q\right)+\sum_{j=2}^{n}\left(\vec{\psi}_{j}\left(p\right)e^{-t_{c}^{-}\left(p\right)\mu_{j}}\sum_{q=1}^{n}\vec{\psi}_{j}\left(q\right)\vec{u}_{0}\left(q\right)\right),\label{eq:infinite time}
\end{equation}

\end{onehalfspace}

here $t_{c}^{-}\left(p\right)$ means the time at which the node $p$
is close to reaching the consensus state. Let $\left\langle \vec{u}_{0}\right\rangle =\frac{1}{n}\sum_{q=1}^{n}\vec{u}_{0}\left(q\right)$
and let us write (\ref{eq:infinite time}) as follows

\begin{equation}
\vec{u}_{t_{c}^{-}}\left(p\right)-\left\langle \vec{u}_{0}\right\rangle =\sum_{j=2}^{n}\left(\vec{\psi}_{j}\left(p\right)e^{-t_{c}^{-}\left(p\right)\mu_{j}}\sum_{q=1}^{n}\vec{\psi}_{j}\left(q\right)\vec{u}_{0}\left(q\right)\right).
\end{equation}

Let us select a node $p$ such that $\vec{\psi}_{2}\left(p\right)$
has the same sign as $\vec{\psi}_{2}\cdot\vec{u}_{0}$. Since $\mu_{2}$
corresponding to $j=2$ is the smallest eigenvalue in the sum on the
right hand side of the expression, this terms tends to 0 slower than
the terms for the other values of $j$. This means that, if we choose
a small enough value of $\delta$, the values of $t_{c}$ and thus
$t_{c}^{-}$ will be very large. Thus, we can ensure that the left
hand side of the equation is small enough that $\sum\limits _{j=3}^{n}\left(\vec{\psi_{j}}(p)e^{-t_{c}^{-}(p)\mu_{j}}(\vec{\psi_{j}}\cdot\vec{u_{0}})\right)<0$.
This implies that

\begin{onehalfspace}
\begin{equation}
\left(\vec{u}_{t_{c}^{-}}\left(p\right)-\left\langle \vec{u}_{0}\right\rangle \right)<\vec{\psi}_{2}\left(p\right)e^{-t_{c}^{-}\left(p\right)\mu_{2}}\left(\vec{\psi}_{2}\cdot\vec{u}_{0}\right).
\end{equation}

Now, because $\left|\vec{u}_{t_{c}^{-}}\left(p\right)-\left\langle \vec{u}_{0}\right\rangle \right|\geq\delta$
we have

\begin{equation}
\delta\leq\left|\vec{u}_{t_{c}^{-}}\left(p\right)-\left\langle \vec{u}_{0}\right\rangle \right|<\left|\vec{\psi}_{2}\left(p\right)e^{-t_{c}^{-}\left(p\right)\mu_{2}}\left(\vec{\psi}_{2}\cdot\vec{u}_{0}\right)\right|.
\end{equation}

Then, the time at which the consensus is reached $t_{c}\left(p\right)$
is bounded by 
\begin{eqnarray}
t_{c}\left(p\right) & \geq t_{c}^{-}\left(p\right)\geq\frac{1}{\mu_{2}} & \ln\left|\frac{\vec{\psi}_{2}\left(p\right)\left(\vec{\psi}_{2}\cdot\vec{u}_{0}\right)}{\delta}\right|.
\end{eqnarray}

\end{onehalfspace}

Finally, the average time of consensus is bounded by

\begin{onehalfspace}
\begin{equation}
\left\langle t_{c}\right\rangle \geq\frac{1}{\mu_{2}n}\sum_{p=1}^{n}\ln\left|\frac{\vec{\psi}_{2}\left(p\right)\left(\vec{\psi}_{2}\cdot\vec{u}_{0}\right)}{\delta}\right|.\label{eq:global time of consensus}
\end{equation}

\end{onehalfspace}

\section{How to Identify Critical Communication Edges?}

The intuition behind the identification of critical edges for consensus
dynamics is very simple. Consensus is a dynamical process in which
information, generically speaking, is transmitted through the nodes
via the edges of the graph. Then, those edges which support most of
the information traffic should be critical for the global agreement
of the network. In other words, the removal of those critical edges---taking
care of not disconnecting the graph---will significantly increase
the average time of consensus of the network. The simplest index fulfilling
this intuition is the \textit{edge betweenness centrality} (EBC) \cite{EBC}.
The EBC of the edge $e$ is defined as

\begin{equation}
EB\left(e\right)=\sum_{v_{i}\in V}\sum_{v_{j}\in V}\dfrac{\rho\left(v_{i},e,v_{j}\right)}{\rho\left(v_{i},v_{j}\right)},
\end{equation}

where $\rho\left(v_{i},e,v_{j}\right)$ is the number of shortest
paths between the nodes $v_{i}$ and $v_{j}$ that go through the
edge $e\in E$, and $\rho\left(v_{i},v_{j}\right)$ is the total number
of shortest paths from $v_{i}$ to $v_{j}$. Obviously, a large value
of the EBC for a given edge indicates that it is critical in the transmission
of information through the network and we should expect that the removal
of that edge increases significantly the average time of consensus
of the network.

Assuming that the information is not only flowing through the shortest
paths allow us to consider a series of other measures that quantify
the amount of potential routes that the information can use to go
from one node to another in the network. The best known of these measures
is the so-called \textit{communicability function} \cite{communicability,communicability review},
which is defined as:

\begin{equation}
G_{uv}=\sum_{k=0}^{\infty}\frac{\left(A^{k}\right)_{uv}}{k!}=\left(e^{A}\right)_{uv}=\sum_{k=1}^{n}e^{\lambda_{k}}\mathbf{q}_{k}(u)\mathbf{q}_{k}(v),\qquad\forall u,v\in V.
\end{equation}

It counts the total number of walks starting at node $u$ and ending
at node $v$, weighting their length by a factor $\frac{1}{k!}$,
hence considering shorter walks more influential than longer ones
(see \cite{communicability,communicability review}).

Here we consider the communicability between a pair of nodes connected
by an edge $\widetilde{G}_{uv}$ where $\left(u,v\right)\in E$. In
this case, it is clear that

\begin{equation}
\widetilde{G}_{uv}=1+\frac{\left(A^{2}\right)_{uv}}{2!}+\frac{\left(A^{3}\right)_{uv}}{3!}+\cdots\qquad\left(u,v\right)\in E.\label{eq:Taylor edge comm-1}
\end{equation}

Then, small values of $\widetilde{G}_{uv}$ indicates that there are
only very long walks that connect the nodes $u$ and $v$ apart from
the edge bounding them together. Because these long walks receive
a large penalization, the edge communicability mainly depends on the
transmission of information through the edge $u,v$.

We now consider a measure that accounts not for the 'volume' of information
transmitted from one node to another in the network but mainly by
the 'quality' of the information transmission. That is, suppose that
two nodes $u$ and $v$ are communicating to each other, the quality
of their communication depends on two factors: (i) how much information
departing from the node $u$ ($v$) arrives at the node $v$ ($u$),
and (ii) how much information departing from the node $u$ ($v$)
returns to that node $u$ ($v$) without ending at its destination.
Then, the goodness of communication increases with the amount of information
which departs from the originator and arrives at its destination,
and decreases with the amount of information which is frustrated due
to the fact that the information returns to its originator without
being delivered to its target. Then, a natural way to account for
this quality of information is by considering the recently proposed
\textit{communicability angle} between a pair of nodes \cite{communicability angles}:

\begin{equation}
\theta_{uv}=\cos^{-1}\frac{G_{uv}}{\sqrt{G_{uu}G_{vv}}}.
\end{equation}

It represents the angle between the position vectors of the nodes
$u$ and $v$ in a Euclidean space, namely a high dimensional Euclidean
sphere where the nodes are placed on the surface separated by their
\textit{communicability distance} \cite{ComDist,ComDist2,ComDist3}: 

\begin{equation}
\xi_{uv}=\sqrt{G_{uu}+G_{vv}-2G_{uv}}.
\end{equation}

The connection between both concepts can be expressed mathematically
as follows:

\begin{equation}
\xi_{uv}^{2}=G_{uu}+G_{vv}-2\sqrt{G_{uu}G_{vv}}\cos\theta_{uv}.
\end{equation}

We notice here that for simple unweighted undirected networks the
communicability angle is bounded as

\begin{equation}
0^{\circ}\leq\theta_{uv}\leq90^{\circ}.
\end{equation}

A large value of communicability between two nodes indicates that
there are many short walks connecting them. In this case the information
has many different routes for going from one node to the other and
there is a kind of redundancy in the topology of the network. Thus,
removing those edges with large communicability is not expected to
have a dramatic effect on the consensus time for this network. On
the contrary, if we remove those edges with poor communicability,
we are removing essential links for the transmission of information
between two nodes due to the fact that very few walks exist that connect
them or they are very long, which will delay significabntly the consensus
process. Extending this reasoning to the communicability angles we
should expect that edges with the largest angles are more probably
the critical ones for the transmission of information in the network.Thus,
the critical edges should be found among those having angles close
to $90^{\circ}.$

\section{Results and Discussion}

\subsection{Critical edges for the time of consensus}

\begin{figure}
\begin{centering}
\includegraphics[width=1\textwidth]{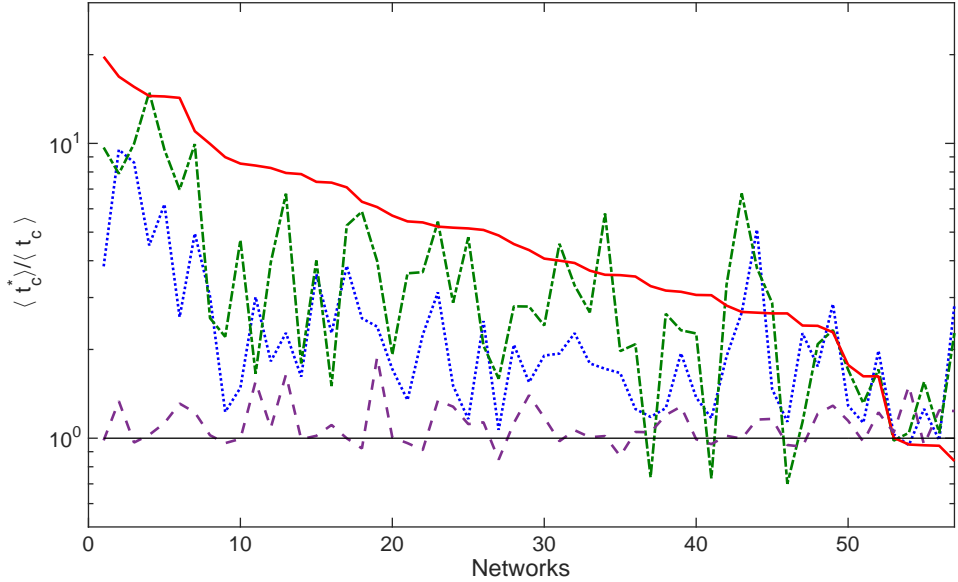} 
\par\end{centering}

\protect\protect\caption{Relative increase in the time of consensus for the real-world networks
studied according to different edge removal strategies. Red solid
line: communicability angle; Green chained line: edge betweenness;
Blue dotted line: Communicability; Magenta spaced broken line: random
removal.}

\label{edge removal results} 
\end{figure}

In order to investigate the role played by the edges on the transmission
of information through the network we design the following experiment.
We consider a series of real-world networks described in the Appendix
which represent complex systems in a variety of scenarios ranging
from social and technological to biomolecular and ecological ones.
We then remove 20\% of their edges by using the following strategies:
(i) removal of the edges with the smallest values of $C_{uv}$; (ii)
removal of the edges with the largest values of $C_{uv}$; random
and independently removal of edges. Here, $C_{uv}$ corresponds to
any of the edge indicators described previously, i.e., edge betweenness
centrality, edge communicability, edge communicability distance and
edge communicability angle. In all cases we take care that the network
does not become disconnected. We then obtain the average time of consensus
for each of the networks generated by using the corresponding removal
strategy and compare them with the original network. The results are
illustrated in the Figure \ref{edge removal results} (see also the
Appendix for specific values), where we give the values of the average
consensus time relative to those of the original networks---values
larger than one indicate relative increase of the consensus time produced
by edge removal respect to the original network.

As can be seen from the Figure \ref{edge removal results} the removal
of 20\% of the edges having the largest communicability angle increases
the average time of consensus by a factor of $5.68\pm4.32$. In other
words, removing the edges with the largest angles multiplies by $5.68$
the time of consensus of the real-world networks studied. In 7 networks
the average time of consensus is increased more than 10 times when
the edges are removed according to the communicability angle, and
in three cases the time of consensus is increased by more than 15
times.

The removal of the edges with the largest EBC also increases significantly
the average time of consensus by a factor of $3.70\pm2.87$. In only
one case the time of consensus is increased by a factor of 10. In
contrast, the random and independent removal of edges increases the
time of consensus only by a factor of $1.12\pm0.20$, and in no case
the time of consensus is duplicated after the random removal of edges.
The removal of the edges ranked according to the other indices studied
here do not increase so significantly the time of consensus of the
networks studied. For instance, removal of edges by the smallest communicability
increases the average time of consensus by $2.41\pm1.70$. 

A significant difference among all the indices studied here is the
fact that removing the edges by the smallest communicability angle
decreases the average time of consensus of the networks. In general,
the decrease of the time of consensus is not very dramatic---as average
it decays by a factor of $0.96\pm0.24$ respect to the original networks,
but in some cases there is an acceleration of the consensus process
by a factor of almost 2. This means that while the largest communicability
angles identify those critical edges whose removal increase significantly
the time of consensus, the smallest angles correspond to such edges
which are redundant in the network and whose elimination in some way
optimize the network for the consensus protocol. Care should be taken
in considering such 'optimization' of the network due to the fact
that removal of these edges could make the networks more vulnerable
to random failures.

So far we have presented the use of the EBC and the communicability
angle based on an intuitive reasoning. As we have seen in the previous
paragraphs this intuition has worked very well due to the fact that
they identify critical edges for consensus dynamics in a very good
way. Here we would like to present some mathematical justification
for these empirical findings which will allow us to better understand
the role of these structural parameters on the dynamical process studied.
We first start with the EBC for which Comellas and Gago \cite{Comellas Gago}
have found the following lower bound. Let $EBC_{max}$ be the maximum
of the EBC in a graph, then

\begin{equation}
EBC_{max}\geq\dfrac{n}{\sqrt{\mu_{2}\left(2\varDelta-\mu_{2}\right)}},
\end{equation}

Consequently, the largest the EBC the smallest the algebraic connectivity,
which implies that the average time of consensus increases according
to (\ref{eq:global time of consensus}). From the structural point
of view this bound is probably telling us that the edges with the
largest EBC are those bottlenecks (or bridges) connecting highly dense
clusters of the network. Indeed, this is what the following bound
obtained by Comellas and Gago indicates \cite{Comellas Gago}:

\begin{equation}
EBC_{max}\geq\dfrac{n}{i\left(G\right)},
\end{equation}

where $i\left(G\right)$ is the isoperimetric number defined as

\begin{equation}
i\left(G\right)=\inf_{S}\dfrac{\left|\partial S\right|}{\left|S\right|},
\end{equation}

where $S$ is a subset of the set of nodes in the network (having
less than the half of the total number of nodes) and $\partial S$
is the set of edges having one endpoint in $S$ and the other in its
complement. Loosely speaking, a large isoperimetric number indicates
that the network does not have structural bottlenecks, i.e., small
sets of edges whose removal disconnect the graph into two almost identical
components. Thus, the relation between EBC and $i\left(G\right)$
indicates that edges with large EBC are contained in networks with
small isoperimetric number, i.e., containing structural bottlenecks.

Let us now turn our analysis to the communicability angle. We first
consider a combined bound for the isoperimetric number obtained by
Mohar \cite{Mohar isoperimetric}:

\begin{equation}
\dfrac{1}{2}\left(\delta-\lambda_{2}\right)\leq i\left(G\right)\leq\sqrt{\Delta^{2}-\lambda_{1}^{2}},
\end{equation}

That is, the isoperimetric number increases with the increase of the
largest eigenvalue $\lambda_{1}$, and with the decrease of the second
largest eigenvalue of the adjacency matrix $\lambda_{2}$. We can
resume this result by saying that the isoperimetric number increases
with the increase of the spectral gap of the adjacency matrix, i.e.,
$\lambda_{1}-\lambda_{2}$. Let us consider what happen to the communicability
angle between a pair of nodes when $\left(\lambda_{1}-\lambda_{2}\right)\rightarrow\infty$.
In this case we have that

\begin{equation}
G_{pq}\rightarrow\psi_{1,p}\psi_{1,q}\exp\left(\lambda_{1}\right),\ \forall p,q\in V.
\end{equation}

Thus, when $\left(\lambda_{1}-\lambda_{2}\right)\rightarrow\infty$
the communicability angle is 

\begin{equation}
\theta_{pq}=\dfrac{G_{pq}}{\sqrt{G_{pp}G_{qq}}}\rightarrow\cos^{-1}1=0^{\circ}.
\end{equation}

In other words, when the graph has large isoperimetric number the
communicability angle tends to zero degrees. A large isoperimetric
constant implies a large algebraic connectivity, i.e., $i\left(G\right)\leq\sqrt{\mu_{2}\left(2\varDelta-\mu_{2}\right)},$
\cite{Mohar isoperimetric}, which indeed implies small average time
of consensus.

\subsection{Time of consensus and global network structure}

Here we investigate how the global structure of networks influences
the average time of consensus. In this case we are guided by the existence
of analytic bounds for the algebraic connectivity of graphs. That
is, the equation (\ref{eq:global time of consensus}) indicates that
the average time of consensus is bounded by the algebraic connectivity
of the network. Thus, we should expect a nice correlation between
these two parameters for networks. However, for a better understanding
of this relation we should dig more deeply about the structural meaning
of the algebraic connectivity. The algebraic connectivity is related
to the minimum degree $\delta$ of a network via the following inequality:

\begin{equation}
\mu_{2}\left(G\right)\leq\dfrac{n\delta}{n-1}.\label{eq:bound min degree}
\end{equation}

By combining two bounds obtained respectively by Alon and Milman \cite{Alon-Milman}
and by Mohar \cite{Mohar} we have that the algebraic connectivity
is bounded as

\begin{equation}
\dfrac{4}{nDiam\left(G\right)}\leq\mu_{2}\left(G\right)\leq\dfrac{8\Delta}{Diam\left(G\right)^{2}}\log_{2}^{2}n.\label{eq:Alon Milman}
\end{equation}

We also consider a lower bound for the algebraic connectivity reported
by Mohar \cite{McKay} in terms of the average path length $\bar{l}\left(G\right)$
of the graph

\begin{equation}
\mu_{2}\left(G\right)\geq\dfrac{4}{2\left(n-1\right)\bar{l}\left(G\right)-\left(n-2\right)}.\label{eq:bound path length}
\end{equation}

These bounds clearly indicate a relation between the average time
of consensus and the metrical properties of the networks.

There are many descriptors used to characterize the structure of graphs
and networks \cite{Estrada_book}. Based on the analytic relations
existing between the time of consensus and some structural parameters
of networks we consider here a few network structural parameters to
be correlated with the average time of consensus of networks. They
include the average node degree

\begin{equation}
\bar{k}\left(G\right)=\frac{1}{n}\sum_{i=1}^{n}k_{i},
\end{equation}

where $k_{i}$ is the degree of the corresponding node and the network
density

\begin{equation}
\varrho\left(G\right)=\frac{\bar{k}}{n-1},
\end{equation}

These parameters can be related to the algebraic connectivity via
the bounds (\ref{eq:bound min degree}) and (\ref{eq:Alon Milman})---we
remind that $\delta\leq\bar{k}\leq\Delta.$

On the other hand we consider the following metrical properties measured
in terms of the shortest-path distance. They are the average shortest
path length, which is given by

\begin{equation}
\bar{d}(G)=\frac{1}{n\left(n-1\right)}\sum_{u,v\in V}d(u,v),
\end{equation}

and the network diameter, which is defined by

\begin{equation}
diam(G)=\max_{u,v\in V(G)}\left\{ d(u,\,v)\right\} .
\end{equation}

We also consider the average distance-sum index---a sort of average
degree based on the sum of distances from a given node to every other
node in the network,

\begin{equation}
s(G)=\dfrac{1}{n}\sum_{u\in V}s\left(u\right),
\end{equation}

where $s\left(u\right)=\sum_{v}d(u,v)$.

\begin{figure}
\protect\includegraphics[width=0.5\textwidth]{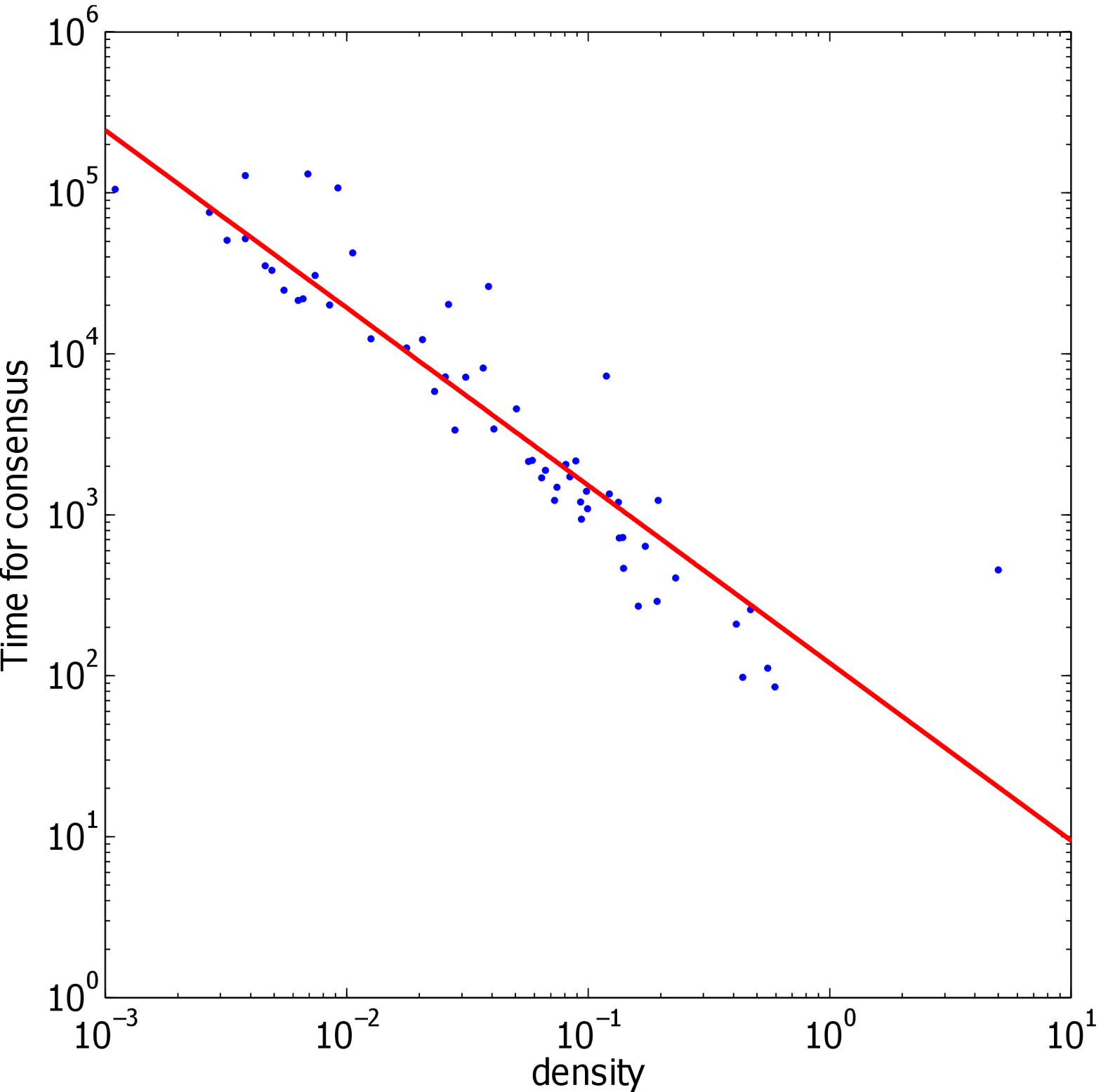}\includegraphics[width=0.5\textwidth]{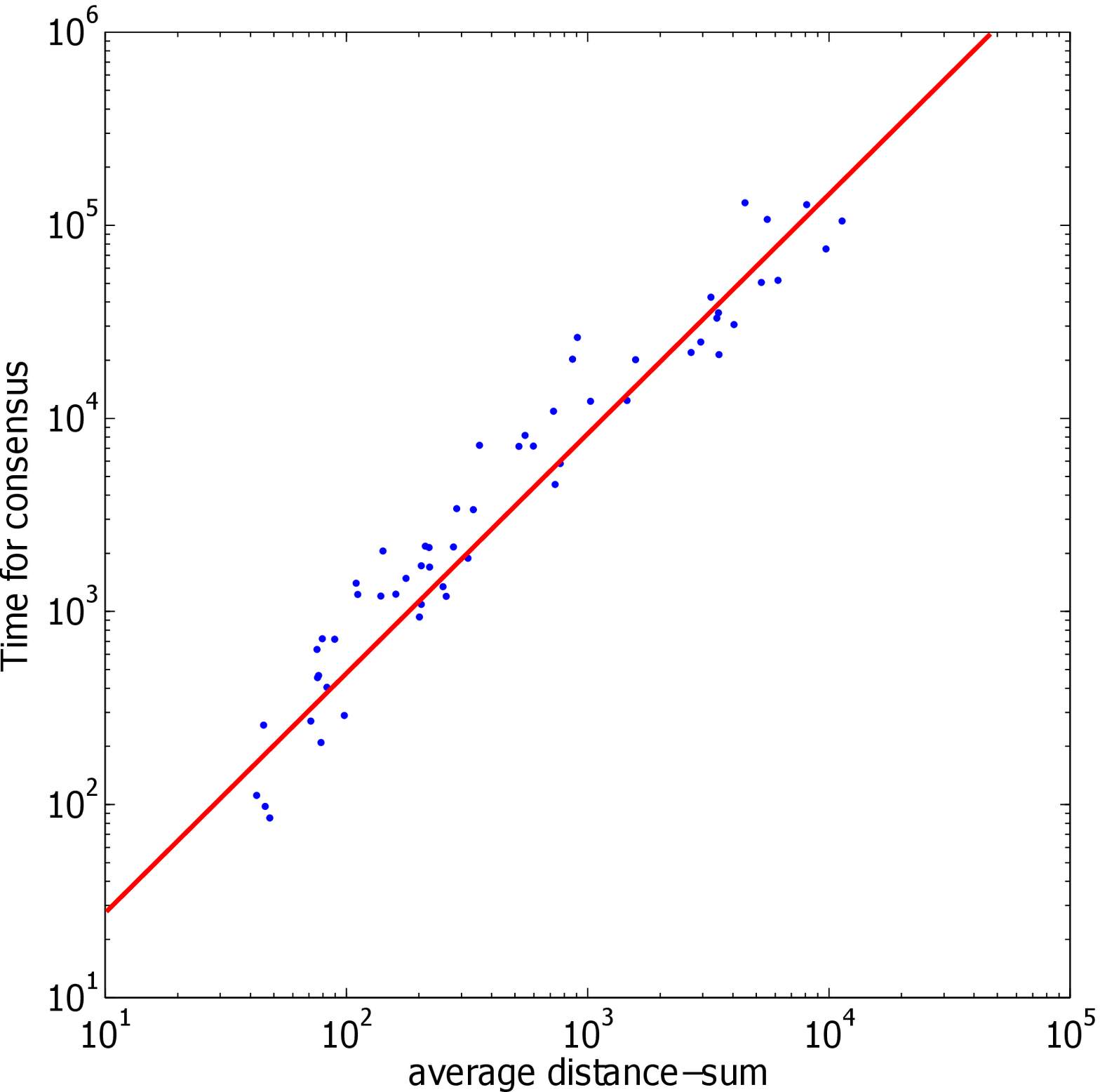}\protect\caption{Illustration of the correlations between the average time of consensus
for the real-world networks studied and the network density and average
distance-sum.}

\label{correlations}
\end{figure}

We then obtain empirical correlations between these measures and the
average time of consensus of all the real-world networks considered
in this work. As expected the algebraic connectivity of the studied
networks display a significant correlation with the average time of
consensus with a Pearson correlation coefficient equal to $-0.792$.
That is, an increase of the algebraic connectivity shorten the time
of consensus of the network as expected from eq. (\ref{eq:global time of consensus}).
However, the Pearson correlation coefficient for the average time
of consensus and the density is $-0.920$ and that with the average
distance-sum is $0.967$ (see Fig. \ref{correlations}), indicating
that these global structural parameters capture much better than the
algebraic connectivity the structural influence over the consensus
dynamics. These results can be understood in the following way. If
we consider networks with the same number of nodes, then the influence
of the algebraic connectivity over the dynamics is significantly larger.
For instance, we have considered all the 11,117 connected graphs with
8 nodes and observed that the Pearson correlation coefficient between
the time of consensus and $\mu_{2}$ is $-0.954$, while those with
the density and average distance sum are $-0.755$ and $0.824$, respectively.
As it can be seen from eq. (\ref{eq:global time of consensus}) the
number of nodes and the Fiedler vector $\vec{\psi}_{2}$ play also
a fundamental role in the determination of the average time of consensus.
Thus, in analyzing the influence of the global structure over the
consensus dynamics, the network density and the average distance-sum
play a more fundamental role than the algebraic connectivity. That
is, increasing the density of the networks and reducing the average
distance-sum of the nodes will decrease significantly the time of
consensus, mainly as a consequence of the fact that information has
significantly more ways to reach the same node from another using
significantly shorter paths.

\section{Conclusions}

We have investigated the relation between local and global structural
parameters over the consensus dynamics on networks. At the local level
we have identified the structural characteristics that make an edge
critical for consensus. That is, those edges whose removal increases
significantly the time necessary for a global consensus in a network.
The removal of edges with the largest edge betweenness centrality,
which accounts for the volume of information flowing through a given
edge, increases the time of consensus in real-world networks by a
factor of $3.70$. On the other hand, the removal of edges based on
their largest communicability angles, which accounts for the quality
of information transmitted through a given edge, increases the consensus
time by a factor $5.68$. We have also considered some global structural
parameters that influence the consensus dynamic of a network. In particular,
the network density and the average distance-sum accounts for more
than $80\%$ of the variance in the time of consensus of a large series
of real-world networks arising in a variety of different scenarios.
In closing, we have identified a few structural parameters---both
local and global---which critically influence the dynamical properties
of networks, allowing further studies to design networks with more
efficient and robust consensus dynamics.

\section*{Acknowledgments}

EE thanks the Royal Society of London for a Wolfson Research Merit
Award. EVE acknowledges CONACYT (Mexico) for a PhD fellowship at the
University of Strathclyde. HA thanks JSPS Grant-in-Aid for Challenging
Exploratory Research No. 15K12137 and by CSTI, Cross-ministerial Strategic
Innovation Promotion Program (SIP), ``Next-generation power electronics''
(NEDO).

\section*{Appendix}
\begin{description}
\item [{Dataset-description}]~
\end{description}
\emph{Biological} \emph{networks}
\begin{itemize}
\item Drosophila PIN: Protein-protein interaction network in \textit{Drosophila
melanogaster} (fruit fly).
\item Hpyroli: Protein-protein interaction network in \textit{H. pyroli}.
\item KSHV: Protein-protein interaction network in \textit{Kaposi sarcoma
herpes virus}.
\item Macaque: The brain network of macaque cortex.
\item Malaria-PIN: Protein-protein interaction network in \textit{P. falciparum}
(malaria parasite). 
\item neurons: neuronal synaptic network of the nematode \textit{C. elegans}.
Included all data except muscle cells and using synaptic connections.
\item PIN-Afulgidus: Protein-protein interaction network in \textit{A. fulgidus}.
\item PIN-Bsubtilis: Protein-protein interaction network in \textit{B. subtilis}.
\item PIN-Ecoli: Protein-protein interaction network in \textit{E. coli}.
\item Transc-yeast: Transcriptional regulation between genes in \textit{Saccaromyces
cerevisiae}.
\item Trans-urchin: Developmental transcription network for sea urchin endomesoderm
development.
\item YeastS: Protein-protein interaction network in \textit{S. cerevisiae}
(yeast).
\end{itemize}
\emph{Ecological} \emph{networks}
\begin{itemize}
\item Benguela: Marine ecosystem of Bengela, off the south-west coast of
South Africa.
\item BridgeBrook: Pelagic species from the largest of set of fifty Adirondack
Lake (NY) food webs.
\item canton: Primarily invertebrates and algae in a tributary, surrounded
by pasture, of the Taieri River in the South Island of New Zealand.
\item Chesapeake: The pelagic portion of an eastern US estuary, with an
emphasis on larger fish.
\item Coachella: Wide range of highly aggregated taxa from the Coachella
Valley desert in Southern California.
\item ElVerde: Insects, spiders, birds, reptiles, and amphibians in a rainforest
in Puerto Rico. 
\item grassland: All vascular plants and all insects and trophic interactions
found inside stems of plants collected from 24 sites distributed within
England and Wales
\item ReefSmall: Caribbean coral reef ecosystem in Puerto Rico/Virgin Island
shelf complex.
\item ScotchBroom: Trophic interactions between the herbivores, parasitoids,
predators, and pathogens associated with broom, \textit{Cytisus scoparius},
collected in Silwood Park, Berkshire, England.
\item Skipwith: Invertebrates in an English pond.
\item StMarks: Mostly macroinvertebrates, fish, and birds associated with
an estuarine seagrass community, \textit{Halodule wrightii}, at the
St. Marks Refuge, Florida, USA.
\item StMartin: Birds and predators and arthropod prey of Anolis lizards
on the island of St. Martin in the northern Lesser Antilles.
\item Stony: Primarily invertebrates and algae in a tributary, surrounded
by pasture, in native tussock habitat, of the Taieri River in the
South Island of New Zealand.
\item Ythan1: Mostly birds, fish, invertebrates, and metazona parasites
in a Scottish estuary.
\item Ythan2: Reduced version of Ythan1, without parasites.
\end{itemize}
\emph{Informational networks}
\begin{itemize}
\item centrality-literature: Citation network of papers published in the
field of Network Centrality.
\item GD: Citation network of papers published in Proceedings of Graph Drawing
during the period 1994-2000.
\item Roget: Vocabulary network of words related by their definitions in
Roget's Thesaurus of the English language. Two words are connected
if one is used in the definition of the other.
\item SmallWorld: Citation network papers which cite Milgram's 1967 Psychology
Today paper or include Small World in the title.
\end{itemize}
\emph{Social} \emph{networks}
\begin{itemize}
\item BF (3, 70, 71): Networks of friendship ties from the communities identified
as 23, 70, and 71 from the Brazilian Farmers longitudinal study on
the adoption of a new corn seed.
\item ColoSpg: The risk network of persons with HIV infection during its
early epidemic phase in Colorado Springs, USA, using analysis of community-wide
HIV/AIDS contact tracing records (sexual partners) during 1985-99.
\item CorporatePeople: American corporate elite formed by the directors
of the 625 largest corporations that reported the compositions of
their boards, selected from the Fortune 1,000 in 1999.
\item dolphins: Social network of a bottlenose dolphins (Tursiops truncates)
population near New Zealand.
\item Drugs: Social network of injecting drug-users (IDUs) who have shared
a needle in the last six months.
\item Galesburg2: Friendship ties among 31 physicians.
\item High-tech: Friendship ties among the employees in a small high-tech
computer firm which sells, installs, and maintains computer systems.
\item hs\_2: Heterosexual contacts, extracted at the Cadham Provincial Laboratory;
a six-month block data from November 1997 to May 1998.
\item Math Method: This network concerns the diffusion of a new mathematics
method in the 1950s. It traces the diffusion of the modern mathematical
method among school systems that combine elementary and secondary
programs in Allegheny County (Pennsylvania, USA.) .
\item PRISON: Social network of inmates in prison who chose ``Which fellows
on the tier are you closest friends with?''
\item Sawmill: Social communication network within a sawmill, where employees
were asked to indicate the frequency with which they discussed work
matters with each of their colleagues.
\item social3: Social network among college students participating in a
course about leadership. The students choose which three members they
want to have on a committee.
\item Zackar: Social network of friendship between members of the Zackary
karate club.
\end{itemize}
\emph{Technological networks}
\begin{itemize}
\item electronic (1-3): Electronic sequential logic circuits parsed from
the ISCAS89 benchmark set, where nodes represent logic gates and flip-flops.
\item Internet-1997: The internet at the Autonomous System (AS) level, as
of September 1997. 
\item Software (Abi, Digital, Mysql, VTK, XMMS): Software network development
for different programs.
\item USAir97: Airport transportation network between airports in the US
in 1997.
\end{itemize}
\begin{table}[H]
\begin{tabular}{llc>{\centering}p{1.5cm}>{\centering}p{1.5cm}>{\centering}p{1.5cm}>{\centering}p{1.5cm}>{\centering}p{1.5cm}}
\hline 
\multicolumn{2}{l}{} &  & \multicolumn{4}{c}{Relative increase of $\left\langle t_{c}\right\rangle $ according
to:} & \tabularnewline
No. & Network & $n$ & $\widetilde{G}_{pq}$ & EBC & $\theta_{pq}$ & Rnd & Ref.\tabularnewline
\hline 
1 & Coachella & 30 & 3.83 & 9.68 & \textbf{19.68} & 0.71 & \cite{Coachella}\tabularnewline
2 & Skipwith & 35 & 9.52 & 7.89 & \textbf{16.85} & 0.76 & \cite{Skipwith}\tabularnewline
3 & electronic3 & 512 & 8.61 & 9.97 & \textbf{15.55} & 1.13 & \cite{electronic}\tabularnewline
4 & Software-XMMS & 971 & 4.51 & \textbf{14.91} & 14.49 & 1.00 & \cite{Softwares}\tabularnewline
5 & electronic2 & 252 & 6.22 & 9.52 & \textbf{14.43} & 0.93 & \cite{electronic}\tabularnewline
6 & hs\_2 & 69 & 2.57 & 6.97 & \textbf{14.27} & 0.95 & \cite{hs_2}\tabularnewline
7 & electronic1 & 122 & 4.95 & 9.95 & \textbf{11.00} & 1.03 & \cite{electronic}\tabularnewline
8 & Software-Mysql & 1480 & 3.03 & 2.55 & \textbf{9.96} & 0.73 & \cite{Softwares}\tabularnewline
9 & centrality-literature & 118 & 1.23 & 2.21 & \textbf{8.98} & 0.81 & \cite{centrality-lit}\tabularnewline
10 & Transc-yeast & 662 & 1.48 & 4.70 & \textbf{8.54} & 1.08 & \cite{transc-yeast}\tabularnewline
11 & social3 & 32 & 3.01 & 1.65 & \textbf{8.41} & 0.95 & \cite{hs_2}\tabularnewline
12 & dolphins & 62 & 1.82 & 3.94 & \textbf{8.25} & 0.79 & \cite{dolphins}\tabularnewline
13 & StMarks & 48 & 2.28 & 6.75 & \textbf{7.94} & 0.85 & \cite{StMarks}\tabularnewline
14 & Drugs & 616 & 1.62 & 1.79 & \textbf{7.87} & 0.97 & \cite{Drugs}\tabularnewline
15 & Software-VTK & 771 & 3.60 & 4.01 & \textbf{7.41} & 0.82 & \cite{Softwares}\tabularnewline
16 & Malaria-PIN & 229 & 2.27 & 1.50 & \textbf{7.36} & 1.23 & \cite{Malaria-PIN}\tabularnewline
17 & CorporatePeople & 1586 & 3.84 & 5.26 & \textbf{7.10} & 1.02 & \cite{Corporate}\tabularnewline
18 & ElVerde & 156 & 2.55 & 5.87 & \textbf{6.34} & 0.77 & \cite{ElVerde}\tabularnewline
19 & Math Method & 30 & 2.39 & 3.99 & \textbf{6.08} & 0.85 & \cite{MathMethod}\tabularnewline
20 & Roget & 994 & 1.72 & 1.92 & \textbf{5.69} & 0.96 & \cite{Roget}\tabularnewline
21 & PINEcoli & 230 & 1.35 & 3.62 & \textbf{5.44} & 0.91 & \cite{PINEcoli-validated}\tabularnewline
22 & Benguela & 29 & 2.24 & 3.66 & \textbf{5.39} & 0.72 & \cite{Skipwith}\tabularnewline
23 & Galesburg2 & 31 & 3.13 & \textbf{5.43} & 5.22 & 0.93 & \cite{Galesburg}\tabularnewline
24 & BridgeBrook & 75 & 1.51 & 2.88 & \textbf{5.18} & 2.17 & \cite{BridgeBrook}\tabularnewline
25 & SmallWorld & 233 & 1.16 & 4.81 & \textbf{5.15} & 1.00 & \cite{SmallW}\tabularnewline
26 & PRISON & 67 & 2.50 & 2.06 & \textbf{5.09} & 1.13 & \cite{Prison}\tabularnewline
27 & Stony & 112 & 1.07 & 1.59 & \textbf{4.87} & 1.01 & \cite{Stony}\tabularnewline
28 & Hi-tech & 33 & 2.08 & 2.80 & \textbf{4.56} & 0.60 & \cite{Hi-Tech}\tabularnewline
29 & Zackar & 34 & 1.55 & 2.80 & \textbf{4.35} & 1.15 & \cite{Zackar}\tabularnewline
30 & Software-Digital & 150 & 1.91 & 2.41 & \textbf{4.07} & 1.33 & \cite{Softwares}\tabularnewline
31 & Trans-Ecoli & 328 & 1.94 & \textbf{4.57} & 4.00 & 0.72 & \cite{transc-yeast}\tabularnewline
32 & GD & 249 & 2.27 & 3.30 & \textbf{3.93} & 0.86 & \cite{SmallW}\tabularnewline
33 & Hpyroli & 710 & 1.79 & 2.67 & \textbf{3.70} & 0.63 & \cite{Hpyroli}\tabularnewline
34 & ScotchBroom & 154 & 1.72 & \textbf{5.78} & 3.58 & 0.98 & \cite{SocthBroom}\tabularnewline
35 & canton & 108 & 1.66 & 1.97 & \textbf{3.57} & 1.01 & \cite{canton}\tabularnewline
36 & Chesapeake & 33 & 1.26 & 2.08 & \textbf{3.54} & 1.55 & \cite{Chesapeake}\tabularnewline
37 & Software-Abi & 1035 & 1.18 & 0.74 & \textbf{3.28} & 0.62 & \cite{Softwares}\tabularnewline
38 & BF-70 & 48 & 1.26 & 2.64 & \textbf{3.17} & 1.40 & \cite{BF}\tabularnewline
39 & Sawmill & 36 & 1.94 & 2.32 & \textbf{3.14} & 1.18 & \cite{Sawmill}\tabularnewline
40 & YeastS & 2224 & 1.37 & 2.27 & \textbf{3.06} & 0.98 & \cite{YeastS}\tabularnewline
41 & Ythan2 & 92 & 1.17 & 0.73 & \textbf{3.06} & 0.96 & \cite{Ythan2}\tabularnewline
42 & PIN-Ecoli & 1251 & 1.90 & \textbf{3.32} & 2.82 & 0.96 & \cite{PINEcoli-validated}\tabularnewline
43 & Internet-1997 & 3015 & 2.67 & \textbf{6.76} & 2.68 & 0.99 & \cite{Internet}\tabularnewline
44 & Macaque & 30 & 5.11 & \textbf{3.78} & 2.67 & 1.04 & \cite{Macaque}\tabularnewline
45 & USAir97 & 332 & 1.47 & \textbf{2.88} & 2.65 & 0.90 & \cite{SmallW}\tabularnewline
46 & Ythan1 & 134 & 1.14 & 0.70 & \textbf{2.65} & 0.97 & \cite{Ythan1}\tabularnewline
47 & neurons-A & 280 & 2.27 & 1.14 & \textbf{2.41} & 0.70 & \cite{neurons}\tabularnewline
48 & grassland-A & 75 & 1.75 & 2.08 & \textbf{2.41} & 0.97 & \cite{grassland}\tabularnewline
49 & BF-71 & 71 & \textbf{2.85} & 2.33 & 2.29 & 0.96 & \cite{BF}\tabularnewline
50 & KSHV & 50 & 1.29 & 1.71 & \textbf{1.77} & 0.79 & \cite{KSHV}\tabularnewline
51 & Trans-urchin & 45 & 1.13 & 1.32 & \textbf{1.62} & 0.78 & \cite{transc-yeast}\tabularnewline
52 & BF-23 & 40 & \textbf{1.98} & 1.71 & 1.62 & 1.11 & \cite{BF}\tabularnewline
53 & ColoSpg & 324 & \textbf{1.03} & 0.98 & 1.00 & 0.99 & \cite{ColoSpg}\tabularnewline
54 & PIN-Afulgidus & 32 & 0.95 & \textbf{1.04} & 0.95 & 0.92 & \cite{PIN-Afulgidus}\tabularnewline
55 & StMartin & 44 & 1.26 & \textbf{1.55} & 0.95 & 0.83 & \cite{StMartin}\tabularnewline
56 & Pin-Bsubtilis & 84 & 0.98 & \textbf{1.04} & 0.94 & 0.96 & \cite{Pin-Bsubtilis}\tabularnewline
57 & ReefSmall & 50 & \textbf{2.84} & 2.28 & 0.84 & 0.79 & \cite{ReefSmall}\tabularnewline
\hline 
\end{tabular}

\protect\caption{Dataset: $n$ number of nodes, $\widetilde{G}_{pq}$ average communicability,
EBC edge-betweenness centrality, $\theta_{pq}$ communicability angle,
Rnd random. The largest increase in the average time of consensus
are boldfaced. }
\end{table}

\end{document}